\documentclass[epjST]{svjour}
\usepackage[T1]{fontenc}
\usepackage[latin9]{inputenc}
\usepackage{mathrsfs}
\usepackage{amsmath}
\usepackage{amssymb}

\usepackage{graphicx}
\usepackage{graphicx,epstopdf}
\def\ie{i.e.~}
\def\etl{$et ~al.$~}
\def\etc{ etc.~}
\begin{document}
\title{The dynamics of co- and counter rotating coupled spherical pendulums}
\author{Blazej Witkowski\inst{1} \and Przemyslaw Perlikowski\inst{1} \fnmsep\thanks{\email{przemyslaw.perlikowski@p.lodz.pl}} \and Awadhesh Prasad\inst{2} \and Tomasz Kapitaniak\inst{1}}
\institute{Division of Dynamics, Technical University of Lodz, 90-924 Lodz, Stefanowskiego 1/15, Poland \and Department of Physics and Astrophysics, University of Delhi, Delhi 110007, India}
\abstract{
The dynamics of co- and counter-rotating coupled spherical pendulums
(two lower pendulums are mounted at the end of the upper pendulum)
is considered. Linear mode analysis shows the existence of three rotating
 modes. Starting from linear modes allow we calculate the  nonlinear normal
modes, which are  and present them in  frequency-energy plots. With the increase
of energy in one mode we observe a symmetry breaking pitchfork bifurcation.
In the second part of the paper we consider energy transfer between
pendulums having different energies. The results for
 co-rotating  (all pendulums rotate in the same
direction) and counter-rotating motion (one of lower pendulums rotates
in the opposite direction) are presented. In general, the energy fluctuations
in counter-rotating pendulums are found to be higher than in the co-rotating case.
} 
\maketitle
\section{Introduction}
\label{intro}
Coupled oscillators can exhibit complex phenomena such as: energy
flows, synchronization, beating, internal resonances, amplitude death,
chaotic and quasiperiodic transients \etc \cite{Pikovsky2001,Strogatz2000,Kuramoto1984,Sabarathinam20133098,Yanchuk2005c,vakakis2008nonlinear,Saxena2012205,Wierschem2012}.
There are numerous studies on the dynamics of the single pendulum and
 coupled pendulums, but mostly devoted to in-plane oscillations \cite{doi:10.1142/S0218127412501003,czolczynski:023129,Brzeski20125347}. In
our studies we investigate the behavior of the coupled spherical pendulums. The
first description of spherical pendulum dynamics has been presented
by Olssen \cite{olsson1978,olsson1981}, who derived the equations of
motion and solved them analytically using Lindstedt-Poincare method.
The obtained solution shows periodic rotation of pendulum for small
but finite displacements. The spherical pendulum is often taken as
a model in quantum mechanics, for example in Refs.  \cite{14,15} where the authors consider
a Hamiltonian system showing its asymptotic properties. In Ref. \cite{Leyendecker2004} the
spherical pendulum is taken as an example to show different ways
of solving Hamiltonian system with holonomic constrains. The authors show
that the Penalty Method can compete with the Lagrange Multiplier Method
and the choice of the method depends on the complexity of the problem and its
expected accuracy. The dynamics of spherical pendulum for increasing
value of energy is presented in Ref. \cite{jp9617128} where the authors consider
several energy levels and present, using analytical and numerical
tools, general scenarios of bifurcations. The global nonlinear stable
manifolds of the spherical pendulum hyperbolic equilibrium with closed
loop attitude control are analyzed by Lee \etl \cite{2011arXiv1103.2822L}.
Their investigations have led us to understand the global stabilization properties
of closed loop control systems on nonlinear spaces. The consequence
of symmetry breaking including PT-transition is shown in Ref. \cite{bender:173}.
The spherical pendulum is also used to model an arm carrying a cup
of coffee \cite{PhysRevE.85.046117}.

The dynamics of double pendulum has been considered in Ref. \cite{raey} where the author used the model consisting of two rigid rods with elastic joints
with the force acting parallel to lower pendulum. The detailed stability
analysis based on the center manifold theorem has been considered for
 hanging down position. Then the periodic oscillations under the varying external
force and damping coefficient have been studied. In another work \cite{3} the general theory
of Lagrangian reduction  is applied to the equations of motion to simplify
the problem and obtain the main form of vibrations and its bifurcations.
The symmetric properties of spherical pendulum motion are investigated,
in details, by Chossat and Bou-Rabee \cite{doi:10.1137/040616681}.
When the symmetry is present in the system one can observe a symmetric
quasiperiodic  and chaotic motions separated by heteroclinic connections.
 The analytical investigation including the
analysis of double spherical pendulum topology was conducted by Hu
\etl \cite{11}. Their work has led us to understand the geometric as well as the dynamical properties
of the systems.

In this paper we analyze the rotational motion of three coupled spherical
pendulums. Using the linear approximation we obtain three independent
linear modes of the pendulum's rotation. In each mode the pendulums
rotate in clock-wise direction with different phase shifts and different
amplitudes. Based on these linear modes we estimate nonlinear normal
modes \cite{Shaw199385,kerschen2009nonlinear,peeters2009nonlinear}
 using path-following algorithm. With the increase of 
total energy of the system the frequencies of the  modes also increase. However, for higher energy
level, the symmetry breaking pitchfork bifurcation occurs in first mode.
For  periodic solutions there is no transfer of  energy between the pendulums.
The energy flows among the pendulums for the co-rotating  and counter-rotating rotations are discussed. 

The paper is organized as follows. The considered model of the coupled
spherical pendulum is introduced in Sec. 2. Sec. 3 contains
the derivation of the eigenfrequencies and eigenvectors of the linearized
systems, which let us compute the nonlinear normal modes. 
The  energy flows between the pendulums are discussed in Sec. 4.
 The conclusions of the results are summarized
in Sect. 5.

\section{Model of two coupled spherical pendulums}
\label{sec:1}

\begin{figure}[h]
\centering{}\includegraphics{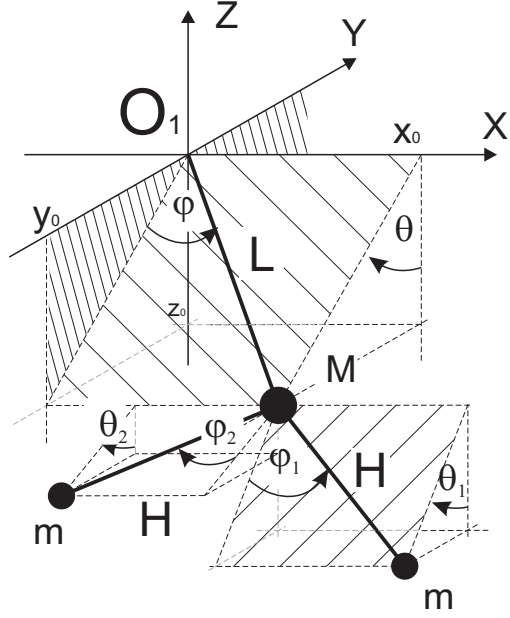}\caption{Schematic figure for the coupled pendulum.}
\label{fig:model}
\end{figure}

We consider the  system  composed of an upper and two lower spherical
pendulums as shown schematically   in Fig. \ref{fig:model}. The upper pendulum
with mass $M$ is suspended on  weightless and inextensible string
of length $L$. At the end of the upper pendulum two identical
pendulums with mass $m$ are suspended on  weightless and inextensible strings
of length $H$. The motion
of upper pendulum is described by two angles $\varphi$ and $\theta$, where $\varphi$ is the angle between
the string $L$ and the plane $YZ$, while angle $\theta$ represents
the angular position of the pendulum around axis $X$. Motions of the
first and the second lower pendulum are described in the same manner
by variables $\varphi_{1}$, $\theta_{1}$ and $\varphi_{2}$,
$\theta_{2}$ respectively. This system is Hamiltonian as there is neither external
 force nor dissipation of energy due to any type of frictional force.

In  Cartesian coordinates the position of the upper pendulum can be written using transformation:
 $x_{0}=L\sin\varphi$, $y_{0}=L\sin\theta\cos\varphi$ and
$z_{0}=-L\cos\theta\cos\varphi$. Similarly, for each lower pendulum the
coordinates can be written as follows: $x_{i}=H\sin\varphi_{i}+x_{0}$,
$y_{i}=H\sin\theta_{i}\cos\varphi_{i}+y_{0}$ and $z_{i}=-H\cos\theta_{i}\cos\varphi_{i}+z_{0}$,
where $i=1,\:2$.
The total kinetic energy is composed of the energy of the upper pendulum and the energy of two lower pendulums. For upper pendulum this energy is given by:
\begin{eqnarray*}
T_{M} & = & \frac{1}{2}M(\dot{x}_{0}^{2}+\dot{y}_{0}^{2}+\dot{z}_{0}^{2})=\frac{1}{2}L^{2}M\left(\cos^{2}\varphi\dot{\theta}^{2}+\dot{\varphi}^{2}\right),
\end{eqnarray*}
while the kinetic energy of the $i-th$ lower pendulum can be expressed
as: 

\begin{eqnarray*}
T_{m_{i}} & = & \frac{1}{2}m(\dot{x}_{i}^{2}+\dot{y}_{i}^{2}+\dot{z}_{i}^{2})=\\
 & = & \frac{1}{2}m\bigg(L^{2}\cos^{2}\varphi\dot{\theta}^{2}+H^{2}\cos^{2}\varphi_{i}\dot{\theta}_{i}^{2}-2HL\cos\varphi_{i}\sin[\theta-\theta_{i}]\sin\varphi\dot{\theta}_{i}\dot{\varphi}+L^{2}\dot{\varphi}^{2}+\\
 &  & +2HL\big(\cos\varphi\cos\varphi_{i}+\cos[\theta-\theta_{i}]\sin\varphi\sin\varphi_{i}\big)\dot{\varphi}\dot{\varphi}_{i}+H^{2}\dot{\varphi}_{i}^{2}+\\
 &  & +2HL\cos\varphi\dot{\theta}\big(\cos[\theta-\theta_{i}]\cos\varphi_{i}\dot{\theta}_{i}+\sin[\theta-\theta_{i}]\sin\varphi_{i}\dot{\varphi}_{i}\big)\bigg),
\end{eqnarray*}

where $i=1,\:2$.
The potential energy of the system is:

\[
V=g\big(L(2m+M)(1-\cos\theta\cos\varphi)+Hm(2-\cos\theta_{1}\cos\varphi_{1}-\cos\theta_{2}\cos\varphi_{2})\big).
\]

The total energy of the system (Hamiltonian) $\mathcal{H}$ is equal
to the sum of kinetic and potential energies of three considered pendulums.
Based on Lagrange equation of the second type one can derive six coupled
second order ODEs. The equations of motion of the upper pendulum are

\begin{eqnarray}
 &  & L\cos\varphi\Big(gM\sin\theta-2LM\sin\varphi\dot{\theta}\dot{\varphi}+LM\cos\varphi\ddot{\theta}+m\sum_{i=1}^{2}\Big(g\sin\theta+H\cos\varphi_{i}\sin[\theta-\theta_{i}]\dot{\theta}_{i}^{2}-\nonumber \\
 &  & 2L\sin\varphi\dot{\theta}\dot{\varphi}-2H\cos[\theta-\theta_{i}]\sin\varphi_{i}\dot{\theta}_{i}\dot{\varphi}_{i}+H\cos\varphi_{i}\sin[\theta-\theta_{i}]\dot{\varphi}_{i}^{2}+L\cos\varphi\ddot{\theta}+\label{eq:tequation}\\
 &  & H\cos[\theta-\theta_{i}]\cos\varphi_{i}\ddot{\theta_{i}}+H\sin[\theta-\theta_{i}]\sin\varphi_{i}\ddot{\varphi_{i}}\Big)=0,\nonumber 
\end{eqnarray}

and

\begin{eqnarray}
 &  & L\Big(gM\cos\theta\sin\varphi+LM\cos\varphi\sin\varphi\dot{\theta}^{2}+LM\ddot{\varphi}+m\sum_{i=1}^{2}\Big(g\cos\theta\sin\varphi+L\cos\varphi\sin\varphi\dot{\theta}^{2}+\nonumber \\
 &  & H\cos[\theta-\theta_{i}]\cos\varphi_{i}\sin\varphi\dot{\varphi}_{i}^{2}-H\cos\varphi\sin\varphi_{i}\dot{\varphi}_{i}^{2}+H\cos[\theta-\theta_{i}]\cos\varphi_{i}\sin\varphi\dot{\theta}_{i}^{2}+\label{eq:pequation}\\
 &  & H\ddot{\varphi_{i}}(\cos\varphi\cos\varphi_{i}+\cos[\theta-\theta_{i}]\sin\varphi\sin\varphi_{i})+2H\sin[\theta-\theta_{i}]\sin\varphi\sin\varphi_{i}\dot{\theta}_{i}\dot{\varphi}_{i}-\nonumber \\
 &  & H\sin[\theta-\theta_{i}]\cos\varphi_{i}\sin\varphi\ddot{\theta_{i}}+H\cos\theta\cos\varphi_{i}\sin\theta_{i}\sin\varphi\ddot{\theta_{i}}+L\ddot{\varphi}\Big)=0.\nonumber 
\end{eqnarray}

The dynamics of the $i-th$ lower pendulum is also described by two
second order ODEs: 
\begin{eqnarray}
 &  & Hm\cos\varphi_{i}\bigg(g\sin\theta_{i}-2H\sin\varphi_{i}\dot{\theta}_{i}\dot{\varphi}_{i}+H\cos\varphi_{i}\ddot{\theta_{i}}-L\Big(\cos\varphi\sin[\theta-\theta_{i}]\dot{\theta}^{2}+\nonumber \\
 &  & 2\cos[\theta-\theta_{i}]\sin\varphi\dot{\theta}\dot{\varphi}+\cos\varphi\big(\sin[\theta-\theta_{i}]\dot{\varphi}^{2}-\cos[\theta-\theta_{i}]\ddot{\theta}\big)+\sin[\theta-\theta_{i}]\sin\varphi\ddot{\varphi}\Big)\bigg)=0,\label{eq:taquation}
\end{eqnarray}
and

\begin{eqnarray}
 &  & Hm\Big(g\cos\theta_{i}\sin\varphi_{i}+L\cos[\theta-\theta_{i}]\cos\varphi\sin\varphi_{i}\dot{\theta}^{2}+H\cos\varphi_{i}\sin\varphi_{i}\dot{\theta}_{i}^{2}-\nonumber \\
 &  & 2L\sin[\theta-\theta_{i}]\sin\varphi\sin\varphi_{i}\dot{\theta}\dot{\varphi}+L\big((-\cos\varphi_{i}\sin\varphi+\cos[\theta-\theta_{i}]\cos\varphi\sin\varphi_{i})\dot{\varphi}^{2}+\label{eq:paquation}\\
 &  & \cos\varphi\sin[\theta-\theta_{i}]\sin\varphi_{i}\ddot{\theta}+(\cos\varphi\cos\varphi_{i}+\cos[\theta-\theta_{i}]\sin\varphi\sin\varphi_{i})\ddot{\varphi}\big)+H\ddot{\varphi_{i}}\Big)=0,\nonumber 
\end{eqnarray}
where $i=1,\:2$. Eqs (1-4) describe the complete dynamics (without simplifications) of the system presented Fig. 1.

\section{Nonlinear Normal Modes of the system}
\label{sec:NNM}

\subsection{Eigenfrequencies}
\label{sec:Eigenfrequencies}

For single spherical pendulum one can observe three modes which
correspond to the periodic solutions. Two of them are rotational modes (symmetrical) and third is a  planar mode where the pendulum swings in a vertical plane. In this paper we
consider only rotational modes. For three coupled pendulums and 
low values of energy one can observe three rotational modes,
each with own eigenfrequency (note that their mirror refection are also present because of symmetry). To calculate these eigenfrequencies we apply the
theory of linear normal modes. Let us assume that the amplitudes of motion of pendulums
are small, hence we can consider that the system performs harmonic
oscillations. To linearize the systems we use the following approximations:
$\sin\mathbf{q}=\mathbf{q}$ and $\cos\mathbf{q}=1$, 
where $\mathbf{q}=\left[\varphi,\theta,\varphi_{1},\theta_{1},\varphi_{2},\theta_{2}\right]^{T}$.
For simplicity we can present the equations of motions, after linearization,  in a  matrix
form: 

\begin{eqnarray}
\mathbf{A}\mathbf{\ddot{q}}+\mathbf{C}\mathbf{q}=0,
\label{eqmatrix}
\end{eqnarray}
where $\mathbf{A}$ and $\mathbf{C}$ are matrices of inertia and stiffness respectively,
and they have got the following forms

\begin{eqnarray}
\mathbf{A}=\left[\begin{matrix}L^{2}(2m+M) & 0 & HLm & 0 & HLm & 0\\
0 & L^{2}(2m+M) & 0 & HLm & 0 & HLm\\
HLm & 0 & H^{2}m & 0 & 0 & 0\\
0 & HLm & 0 & H^{2}m & 0 & 0\\
HLm & 0 & 0 & 0 & H^{2}m & 0\\
0 & HLm & 0 & 0 & 0 & H^{2}m
\end{matrix}\right],\\
\mathbf{C}=\left[\begin{matrix}gL(2m+M) & 0 & 0 & 0 & 0 & 0\\
0 & gL(2m+M) & 0 & 0 & 0 & 0\\
0 & 0 & gHm & 0 & 0 & 0\\
0 & 0 & 0 & gHm & 0 & 0\\
0 & 0 & 0 & 0 & gHm & 0\\
0 & 0 & 0 & 0 & 0 & gHm
\end{matrix}\right].
\end{eqnarray}

For rotational modes, in three dimensional phase space, the periodic
solutions are rotations around the hanging down position. For that reason
we assume the solution of Eq. (\ref{eqmatrix}) as follows: 
\begin{eqnarray}
 &  & \mathbf{q}=\left[\begin{matrix}\varphi\\
\theta\\
\varphi_{1}\\
\theta_{1}\\
\varphi_{2}\\
\theta_{2}
\end{matrix}\right]=\left[\begin{matrix}\Psi_{\varphi}\sin[\omega t]\\
\Psi_{\theta}\cos[\omega t]\\
\Psi_{\varphi_{1}}\sin[\omega t]\\
\Psi_{\theta_{1}}\cos[\omega t]\\
\Psi_{\varphi_{2}}\sin[\omega t]\\
\Psi_{\theta_{2}}\cos[\omega t]
\end{matrix}\right]=\mathbf{\Psi}\left[\begin{matrix}\sin[\omega t]\\
\cos[\omega t]\\
\sin[\omega t]\\
\cos[\omega t]\\
\sin[\omega t]\\
\cos[\omega t]
\end{matrix}\right],\quad\omega>0,\quad t>0,\,\begin{matrix}\forall\\

\end{matrix}\Psi_{j=\{1,2,3\}}\in\mathbb{R}.\label{eqsol}
\end{eqnarray}

For linear oscillations of the system we observe the symmetry in two parallel
planes (XY, YZ), hence $\Psi_{\varphi}=\Psi_{\theta}$, $\Psi_{\varphi_{1}}=\Psi_{\theta_{1}}$,
$\Psi_{\varphi_{2}}=\Psi_{\theta_{2}}$. One can calculate the eigenfrequencies,
$\omega_{1,2,3}$,  for which the periodic solutions are observed in the system, 
using the relation $(\mathbf{C}-\omega^{2}\mathbf{A})\Psi=0$. The determinant $\det|C-\omega^{2}A|$ has to vanish, which  gives three independent frequencies:
\begin{eqnarray}
\omega_{1} & = & \sqrt{\frac{g}{H}},\nonumber \\
\omega_{2,3} & = & \sqrt{\frac{g}{L}}\sqrt{a\pm b},\label{eq:frequency}\\
\nonumber 
\end{eqnarray}

where $a=\frac{(H+L)(2m+M)}{2HM}$ and $b=\frac{\sqrt{2m+M}\sqrt{2(H+L)^{2}m+(H-L)^{2}M}}{2HM}$. 

\begin{figure}[ht]
\centering{}\includegraphics{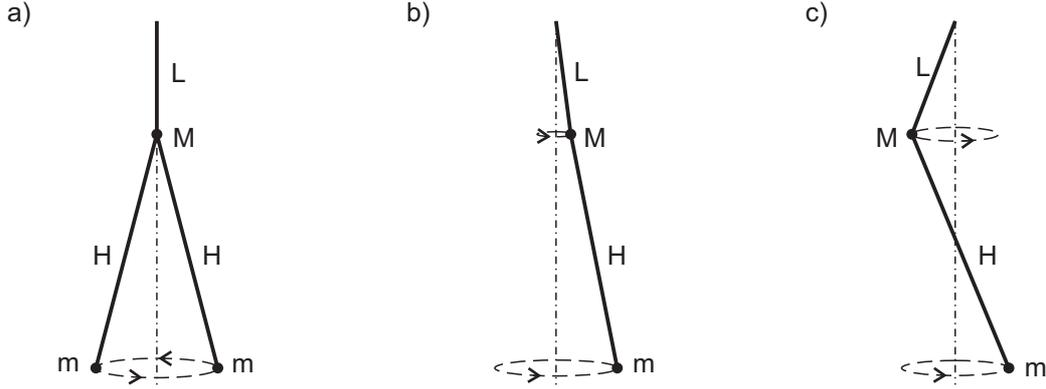}
\caption{The schematic representation of mode shapes for (a) $\omega=\omega_{1}$ (b) $\omega=\omega_{2}$ and
(c) $\omega=\omega_{3}$. }
\label{fig:LNM:freq} 
\end{figure}

The schematic representation of mode shapes corresponding to all three eigenfrequencies 
are shown in Fig. 2. Frequency $\omega_{1}$ corresponds to the
following solution: the phase shift between lower pendulums is $\pi$,
which causes a balance of forces acting on the upper pendulum, hence the
upper pendulum is at rest (see Fig. \ref{fig:LNM:freq} (a)). For
frequency $\omega_{2}$ the upper and lower pendulums rotate in-phase,
lower pendulums have the same amplitudes while the amplitude of upper
pendulum is different (Fig. \ref{fig:LNM:freq} (b)). For frequency
$\omega_{3}$ the phase of upper pendulum is shifted by $\pi$ compared
to the phases of both lower pendulums (Fig. $\ref{fig:LNM:freq}$ (c)).

The eigenvectors, corresponding the eigenfrequencies,  can be written as:
\begin{eqnarray*}
\Psi_{1}=\left[\begin{matrix}0,0,\Psi_{\theta_{1}},\Psi_{\theta_{1}},-\Psi_{\theta_{1}},-\Psi_{\theta_{1}}\end{matrix}\right]^{T},
\end{eqnarray*}

\begin{eqnarray}
\Psi_{2,3}=\left[\begin{matrix}\Psi_{\theta},\Psi_{\theta},\Psi_{\theta}(c\mp d),\Psi_{\theta}(c\mp d),\Psi_{\theta}(c\mp d),\Psi_{\theta}(c\mp d)\end{matrix}\right],
\end{eqnarray}

where $c=\frac{(H-L)(2m+M)}{4Hm}$ and $d=\frac{\sqrt{2m+M}\sqrt{2(H+L)^{2}m+(H-L)^{2}M}}{4Hm}$.

\subsection{Nonlinear normal modes}
\label{sub:NNMspace}

\begin{figure}[ht]
\centering{}\includegraphics{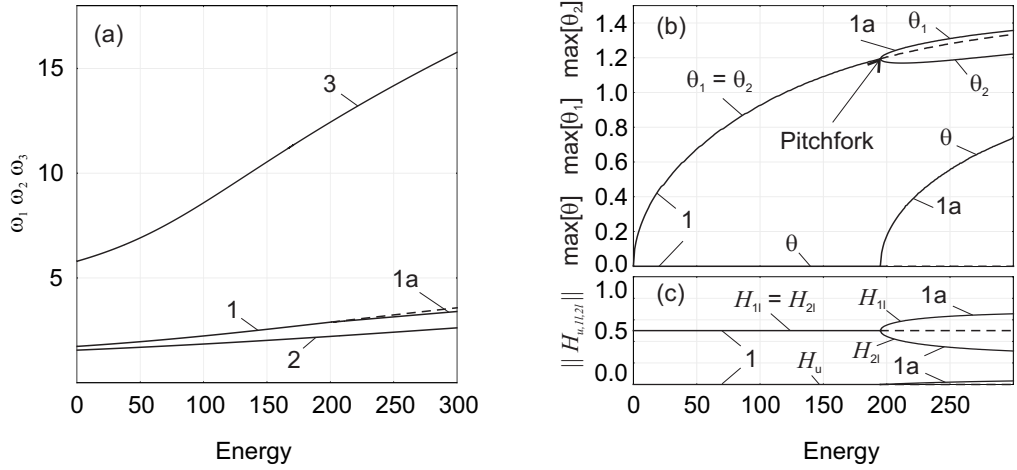}
\caption{ (a) Frequency--energy plot of the investigated nonlinear system. Each branch
begins with frequency calculated for linear system (No. 1 - $\omega_{1}$,
No. 2 - $\omega_{2}$ and No. 3 - $\omega_{3}$). (b) The 
 maximum amplitudes of  solutions  as a function of energy 
for branches No. 1 and No. 1a. At $\mathcal{H} \sim 197$
we observe a pitchfork bifurcation which breaks the symmetry in the system,
and hence the upper pendulum starts to oscillate ($\theta\neq0$) and
amplitudes of lower pendulums ($\theta_{1}$ and $\theta_{2}$) have
different values. (c) The  normalized energies ($\left\Vert \mathcal{H}\right\Vert =1$)
for each pendulum as a function of  total energy of the system for branches No.
1 and No. 1a. The solid and dashed lines correspond to respectively stable
and unstable periodic solutions.}
\label{fig:NNM:rotation:freq-1}
\end{figure}

In numerical calculations we assume the following values of system parameters:
$M=2$, $m=2$, $L=1$, and $H=3$.  After substituting these values in  Eqs. \eqref{eq:frequency}
we obtain the following linear eigenfrequencies: $\omega_{1}=1.732$,
$\omega_{2}=1.553$ and $\omega_{3}=5.7959$. Each of them describes the periodic solution of the linearized system. When we have applied one of them to the nonlinear system (Eqs $(\ref{eq:tequation},\ref{eq:pequation},\ref{eq:taquation},\ref{eq:paquation})$), even assuming small
energy,  
we obtain a quasiperiodic orbit (KAM tori) which is located close to
the  periodic solution of the linearized system. To correct the obtained solution,
 we have applied the Newton-Raphson algorithm. The 
integration of ODEs are performed with  Runge-Kutta-Fehlberg (4,5) method.
 In Figure $\ref{fig:NNM:rotation:freq-1}$(a)
we present, on the frequency - energy plot, how the frequencies of
three periodic solutions  change with the increase of the total
energy $\mathcal{H}$. Each branch is calculated in the following
way: for starting point ($\mathcal{H\approx}0.02$) we take initial
conditions according to linear eigenvector, then we correct the obtained
solution by Newton-Raphson scheme to periodic orbit. In  next step
this solution is perturbed (we add $1\%$ of current total energy)
so the energy is shifted to higher level and once again the correction
is applied. The described procedure is repeated until the energy of the system
reaches $\mathcal{H}=300$. The successive increase of total energy
$\mathcal{H}$ causes the increase of three frequencies, hence the periods
of rotations become shorter. The maximum value of energy can is
 $\mathcal{H}=300$ because for that value the system has singular point - the amplitude of upper pendulum reaches $\pi/2$.
Therefore for branch No. 3 the  maximum amplitude is close to $\pi/2$.
Further increase of the energy ($\mathcal{H}>300$) for branches No.
1, 1a and 2 causes only the increase of the amplitude and the rotational velocity,
but no new phenomena are observed.

Branch No. 1 corresponds to clock-wise rotations of lower pendulums
with phase shifted by $\pi$ while the upper one remains in hanging down position (see Fig.
\ref{fig:LNM:freq}(a)). For energy level equal
to $\mathcal{H}=197$ the symmetry breaking pitchfork bifurcation occurs.
It is indicated by the appearance of No. 1a branch. After the bifurcation the 
frequencies of two solutions stay close in the whole range of considered
energy. The changes of the maximum amplitudes of solutions
along branches No. 1 and 1a are shown in  Fig. \ref{fig:NNM:rotation:freq-1}(b).
The solid  and dashed lines correspond respectively to stable and unstable
periodic solutions. For branch
No. 1a, the oscillations of lower pendulums are asymmetric this causes that upper pendulum starts to oscillate increasing its amplitude with growing energy. The amount of energy in each pendulum for branches
No. 1 and No. 1a is presented in Fig. \ref{fig:NNM:rotation:freq-1}(c).
We normalize the total energy to one ($\left\Vert \mathcal{H}\right\Vert =1$)
and show its participation in each pendulum. It is easy to see that
for branch No. 1 the energy is equally distributed between lower pendulums.
The upper pendulum is not moving before the bifurcation, hence its energy is equal to zero. 
For branch No. 1a the energy of first lower pendulum (energy $\mathcal{H}_{1l}$) starts to increase
while the energy of the second one decreases (energy $\mathcal{H}_{2l}$). The energy of the upper
pendulum also increaes after the bifurcation.

\begin{figure}[ht]
\centering{}\includegraphics{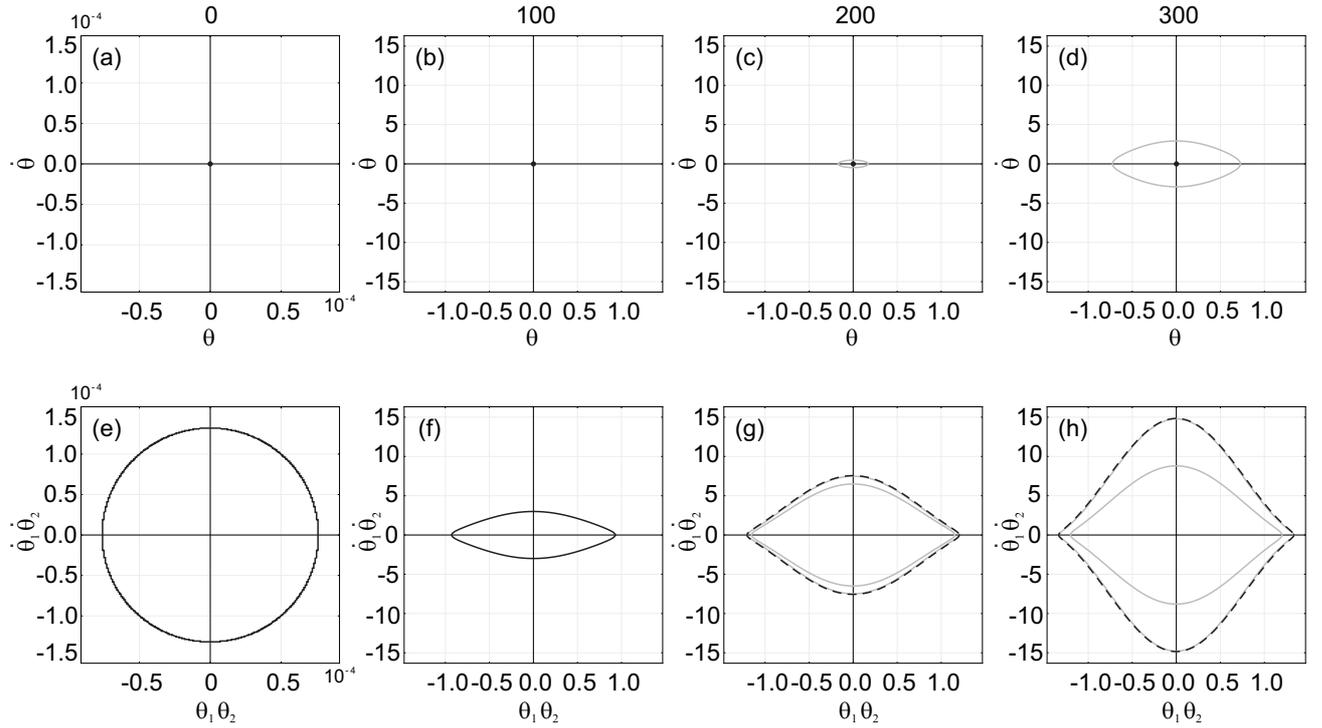}
\caption{The phase space trajectories of nonlinear normal modes along branches No. 1 and No. 1a.
The total energy $\mathcal{H}$ is increased from left to right: $\mathcal{H}=0.02$
(a,e), $\mathcal{H=}100$ (b,f), $\mathcal{H}=200$ (c,g) and $\mathcal{H}=300$
(d,h). The upper and lower rows show  the motions of upper and lower pendulums 
respectively in phase space.  The solid (stable
solutions) and dashed (unstable solutions) black lines show periodic
orbits along branch No. 1 while grey lines present the stable orbits
along branch No. 1a.}
\label{fig:NNM:rotation:1-1}
\end{figure}

The change of the shape of periodic orbits for  branches No. 1 (black
line) and No. 1a (grey line) are shown in Fig. \ref{fig:NNM:rotation:1-1}.
In upper and lower rows, the trajectories of upper pendulum and
lower pendulums are presented respectively. For branch No. 1 the low
energies solutions are nearly harmonic however for higher values of $\mathcal{H}$
we observe  the deformation around the maximum amplitude of pendulums.
Therefore, the time when the pendulums are barely moving becomes longer
in comparison to the rest period of oscillations. The periodic solution
loses stability in the pitchfork bifurcation (the continuous line changes
to the dashed one) and the asymmetric orbits appear (branch No. 1a). 

\begin{figure}[ht]
\centering{}\includegraphics{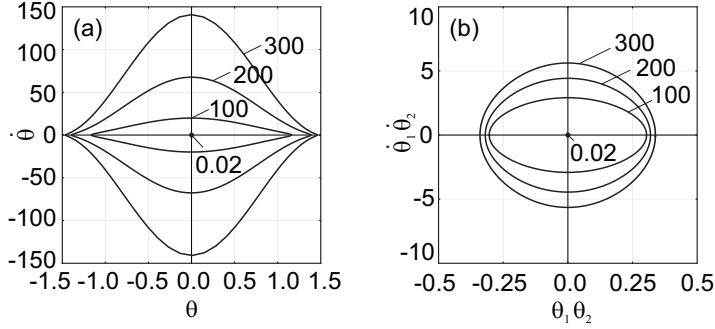}
\caption{ The  projections of trajectories in  phase space of nonlinear normal modes of branch No. 3 
for different  energies: $\mathcal{H}=0.02$, $\mathcal{H}=100$, $\mathcal{H}=200$ and $\mathcal{H}=300$ of
(a) upper and (b) lower pendulums.}
\label{fig:NNM:rotation:3-1}
\end{figure}

For branch No. 2, corresponding to three pendulums oscillating in-phase,
we observe a slow increase of oscillation amplitude with the increase of
total energy. The periodic solutions along branch No. 2 change
their shapes similar to branch No. 1 (not presented).
 For the third branch (No. 3) the upper pendulum rotates in anti-phase
to lower pendulums. The changes of shape of periodic solutions are
presented in Fig. \ref{fig:NNM:rotation:3-1}. It is easy to see that,
for low energy level,  the orbits are nearly harmonic, while for higher
levels of energy they become non-harmonic with visible nonlinear effects
around the maximum amplitude. For $\mathcal{H}=300$ the maximum amplitude
of upper pendulum reaches the singular point $\theta=\pi/2$ (see Fig.
\ref{fig:NNM:rotation:3-1}(a)). In the case of branches No. 1 and No. 2 the amplitudes
of lower pendulums increase much faster than the amplitude of the upper
one, while for branch No. 3 we observe an opposite behavior.

\section{Energy transfer among pendulums}
\label{sec:energytransfer}

In this section we present the energy transfer between lower pendulums
via the upper pendulum. In the case of periodic solutions there is no
energy transfer, \ie, the energy of each pendulum is constant.
To observe the exchange of energy between the pendulums one has to perturb the
periodic motion. This is done by introducing a  small mismatch in the initial conditions.
For low energy this leads to appearance of quasi-periodic motion (KAM tori) while for larger
perturbations one can observe the chaotic behavior. Additionally,
in this section we take into account  counter-rotating solutions.
To achieve these   we change the sign of the initial velocities ($\dot{\varphi}_{2}$
and $\dot{\theta}_{2}$) of one of lower pendulums, \ie,  forcing one pendulum to move
in the opposite direction. As discussed below, we observe that all counter-rotating solutions are chaotic.

\subsection{Perturbation to first mode,  $\omega_{1}$}
\begin{figure}[ht]
\noindent \centering{}\includegraphics{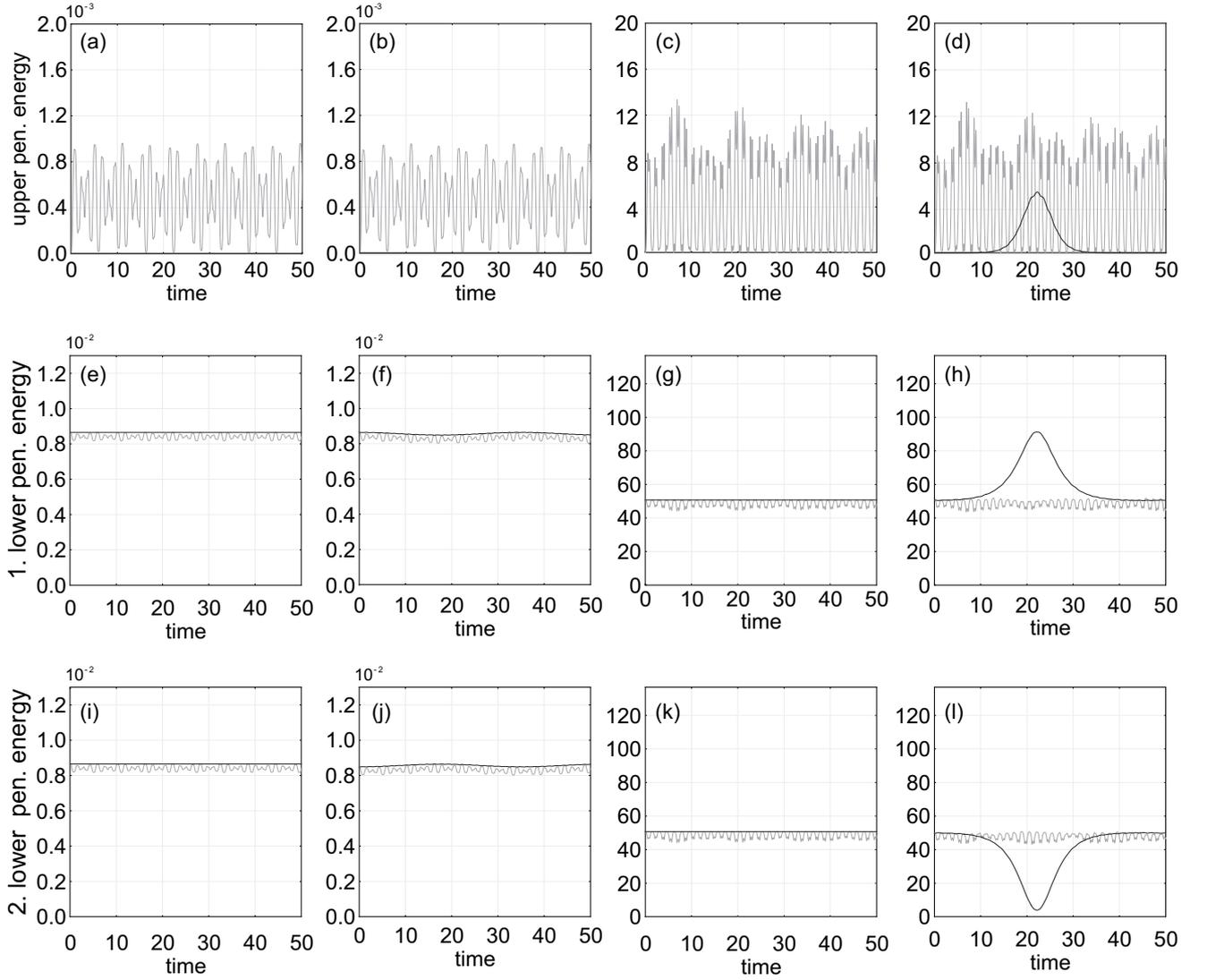}
\caption{ The variation of energies as a function of time for 
the first mode, $\omega_{1}$. The figures of  upper row (a-d) correspond to the upper pendulum
 while the lower two rows (e-l) are for lower pendulums.  The  black and grey lines in each figure represent
 co- and counter rotating solutions respectively.  The trajectories for  (a, e, i) low  energy
level with identical initial conditions, (b, f, j) low  energy
level with mismatched initial conditions, (c, g, k) high  energy
level ($\mathcal{H}=100$) with identical initial conditions, and (d,
h, l) high energy level ($\mathcal{H}=100$) with mismatched
initial conditions.}
\label{fig:LNM:I:en} 
\end{figure}

For low level of the total energy ($\mathcal{H}=0.017035$) we take the
following initial conditions: $\theta=0.0$, $\varphi=0.0$, $\theta_{1}=0.012656702$,
$\varphi_{1}=0.0$, $\theta_{2}=-0.012656702$, $\varphi_{2}=0.0$,
$\dot{\theta}=0.0$, $\dot{\varphi}=0.0$, $\dot{\theta}_{1}=0.0$,
$\dot{\varphi}_{1}=0.02192759$, $\dot{\theta}_{2}=0.0$, $\dot{\varphi}_{2}=-0.02192759$,
which correspond to $1\%$ perturbation of periodic solution. For
counter-rotating the velocity of the second lower pendulum is equal to
$\dot{\varphi}_{2}=0.02192759$ (the velocity $\dot{\theta}_{2}$
is equal to zero).
In Fig. $\ref{fig:LNM:I:en}$(a, e, i) we present how the energies
of upper and lower pendulums change in time for co-rotating (black line) 
and counter-rotating (grey line)
without additional perturbation. For co-rotating motion the pendulums
do not transfer the energy between each other because of static upper pendulum, and
 synchronized periodic motions of lower pendulums.  For counter-rotating motion
the lower pendulums transfer energy via  upper pendulum and motions are chaotic.
For higher energy, after addition of  small perturbation ($\delta\theta_{1}=0.01\theta_{1}$
and $\delta\dot{\varphi}_{1}=0.01\dot{\varphi}_{1}$)  trajectories are shown 
in Fig. $\ref{fig:LNM:I:en}$(b, f, j). In both cases, either co- nor counter rotations, lower pendulums exchange energy via the upper pendulum.
The co-rotating solution does not remain periodic  but it becomes quasi-periodic. In quasi-periodic motion one can distinguish two independent frequencies - one of them remains from periodic motion the second one is responsible for modulation. This second frequency is much
smaller than the frequency of original periodic orbit. This cause that we observe
a slow transfer of energy. However the counter-rotating motion
 remains  chaotic. Note that the
energy is transferred with the same frequency but the amplitude of upper
pendulum's motion is much higher for counter-rotating than for co-rotating.

In Fig. $\ref{fig:LNM:I:en}$(c,g,k) we show the trajectories for high energy level,
 $\left(\mathcal{H}=101.468\right)$ with the following 
initial conditions: $\theta=0.0$, $\varphi=0.0$, $\theta_{1}=0.92996374$,
$\varphi_{1}=0.0$, $\theta_{2}=-0.92996374$, $\varphi_{2}=0.0$,
$\dot{\theta}=0.0$, $\dot{\varphi}=0.0$, $\dot{\theta}_{1}=0.0$,
$\dot{\varphi}_{1}=1.79562871$, $\dot{\theta}_{2}=0.0$, $\dot{\varphi}_{2}=-1.79562871$. 
For counter-rotating case initial
velocity of second pendulum is changed to $\dot{\varphi}_{2}=1.79562871$.
The system behaves in similar way as in the low energy level cases.
The only difference is in the amplitude of motion. However, for sufficiently high energy level, at
$ \left(\mathcal{H}=100 \right)$, as shown in
Fig. $\ref{fig:LNM:I:en}$(d, h, l), we observe higher 
energy transfer between pendulums for co-rotating case.

\subsection{Perturbation to second  mode,  $\omega_{2}$}

\begin{figure}[ht]
\centering{}\includegraphics{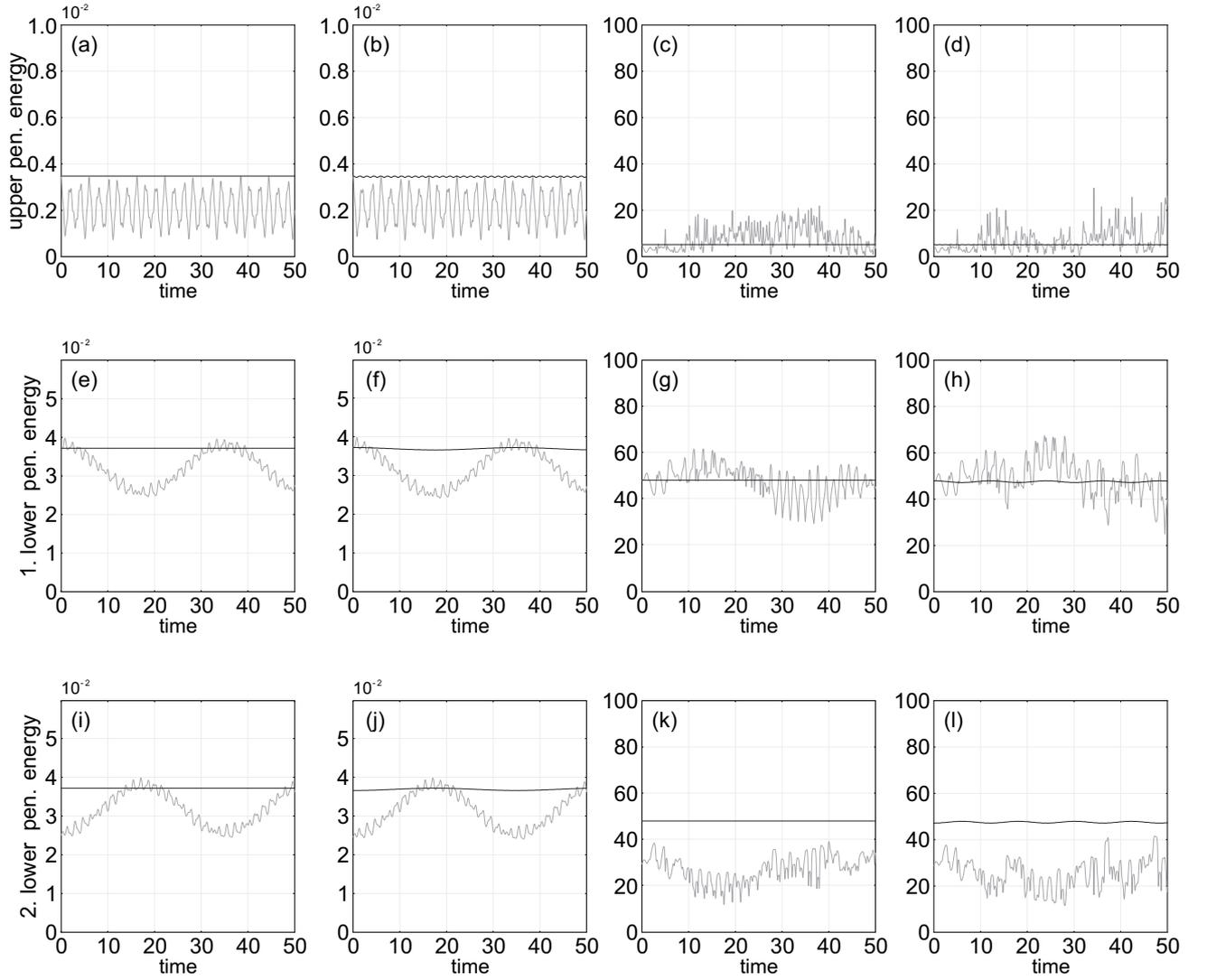}
\caption{ The variation of energies as a function of time for
the second mode, $\omega_{2}$. The figures of  upper row (a-d) correspond to the upper pendulum
 while the lower two rows (e-l) are for lower pendulums.  The  black and grey lines in each figure represent
 co- and counter rotating solutions respectively.  The trajectories for
(a, e, i) low  energy level with
identical initial conditions, (b, f, j) low energy level with
mismatched initial conditions, (c, g, k) high  energy level ($\mathcal{H}=100.989$)
with identical initial conditions, and (d, h, l) high  energy level
($\mathcal{H}=100.989$) with mismatched initial conditions.}
\label{fig:LNM:II:en} 
\end{figure}

For low energy level ($\mathcal{H}=0.0778345$) the following initial
conditions are taken: $\theta=0.01745387$, $\varphi=0.0$, $\theta_{1}=0.02384009$,
$\varphi_{1}=0.0$, $\theta_{2}=0.02384009$, $\varphi_{2}=0.0$,
$\dot{\theta}=0.0$, $\dot{\varphi}=0.02710668$, $\dot{\theta}_{1}=0.0$,
$\dot{\varphi}_{1}=0.03702315$, $\dot{\theta}_{2}=0.0$, $\dot{\varphi}_{2}=0.03702315$.
For counter-rotating solution the initial velocity of second pendulum is
$\dot{\varphi}_{2}=-0.03702315$ (the velocity $\dot{\theta}_{2}$
is equal to zero) is considered.
In  Fig. $\ref{fig:LNM:II:en}$(a, e, i) we show the  change of energies as a function of time
 for periodic motion for co-rotating and chaotic oscillations
for counter-rotating. For co-rotating case the pendulums do not transfer the 
energy between each other. The upper pendulum is moving in-phase with
lower pendulums. When counter-rotating motion is observed lower pendulums
transfer energy to upper pendulum and vice verse. The fluctuation in 
energy indicates that motions are chaotic. 

Next we add the following perturbation to initial conditions: $\delta\theta_{1}=0.01\theta_{1}$
and $\delta\dot{\varphi}_{1}=0.01\dot{\varphi}_{1}$. In Fig. $\ref{fig:LNM:II:en}$(b,
f, j) the change of energy in time for co-rotating and counter-rotating
motions are presented. In both these cases lower pendulums exchange energy
via the upper one. Energy is transferred with the similar frequency
but the amount of the  exchanged energy is much bigger for counter-rotating case
than for co-rotating.

For higher energy level $\left(\mathcal{H}=100.989\right)$  the following
initial conditions are considered: $\theta=0.66836221$, $\varphi=0.0$,
$\theta_{1}=0.817659$, $\varphi_{1}=0.0$, $\theta_{2}=0.817659$,
$\varphi_{2}=0.0$, $\dot{\theta}=0.0$, $\dot{\varphi}=1.14578007$,
$\dot{\theta}_{1}=0.0$, $\dot{\varphi}_{1}=1.3488661$, $\dot{\theta}_{2}=0.0$,
$\dot{\varphi}_{2}=1.3488661$. The trajectories are shown  (Fig. $\ref{fig:LNM:II:en}$(c, g, k)). 
For counter-rotating case the initial velocity of the second pendulum
is  changed to $\dot{\varphi}_{2}=-1.3488661$. Counter-rotating solution continues to oscillate  chaotically (Fig. $\ref{fig:LNM:II:en}$(d, h, l)). 
In the co-rotating case lower pendulums oscillate in quasiperiodic
way. However, for higher  energy level ($\left(\mathcal{H}=100.989\right)$), the period of the energy transfer is much shorter than for lower energy level.

\subsection{Perturbation to third mode, $\omega_{3}$}
\begin{figure}[ht]
\centering{}\includegraphics{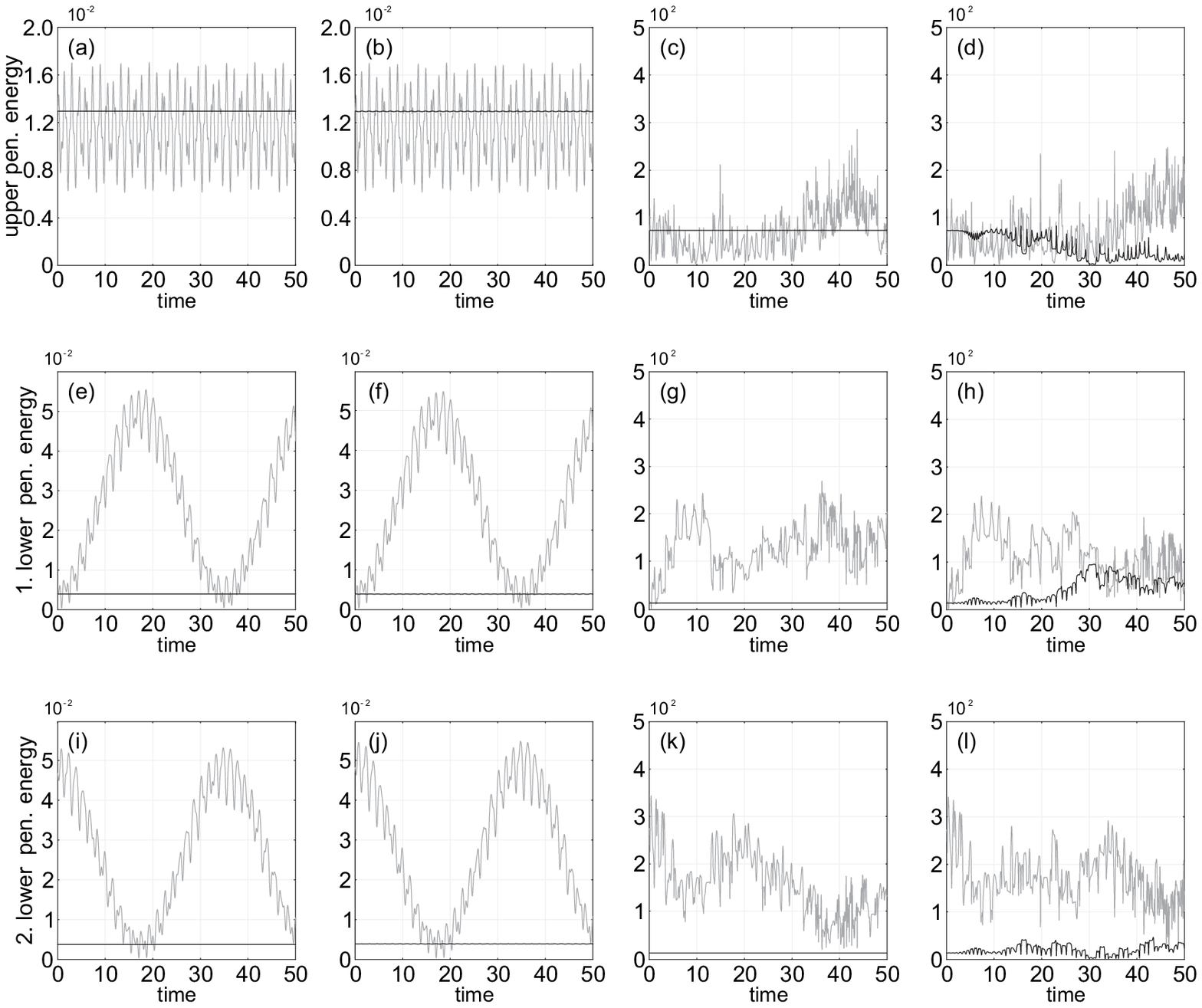}
\caption{ The variation of energies as a function of time for
the third mode, $\omega_{3}$. The figures of  upper row (a-d) correspond to the upper pendulum
 while the lower two rows (e-l) are for lower pendulums.  The  black and grey lines in each figure represent
 co- and counter rotating solutions respectively.  The trajectories for
(a, e, i) low  energy level with
identical initial conditions, (b, f, j) low  energy level with
mismatched initial conditions, (c, g, k) high  energy level ($\mathcal{H}=100.898$)
with identical initial conditions, and (d, h, l) high  energy level
($\mathcal{H}=100.898$) with mismatched initial conditions.}
\label{fig:LNM:III:en}
\end{figure}

Now we consider the third mode where upper and lower pendulums move in
opposite directions. For low energy level ($\mathcal{H}=0.0208567$)
with the following initial conditions $\theta=0.017453269$,
$\varphi=0.0$, $\theta_{1}=-0.0063880017$, $\varphi_{1}=0.0$, $\theta_{2}=-0.0063880017$,
$\varphi_{2}=0.0$, $\dot{\theta}=0.0$, $\dot{\varphi}=0.10115284$,
$\dot{\theta}_{1}=0.0$, $\dot{\varphi}_{1}=-0.037024141$, $\dot{\theta}_{2}=0.0$,
$\dot{\varphi}_{2}=-0.037024141$ are taken. For counter-rotating solution we assume
$\dot{\varphi}_{2}=0.037024141$ (Fig. $\ref{fig:LNM:III:en}$ (a, e, i)) 
Co-rotating solution is periodic and pendulums do not exchange energy
between each other. In counter-rotating case we can observe a chaotic
beating. Similar behavior is observed for  higher energy level $\left(\mathcal{H}=100.898\right)$
as shown in Fig. $\ref{fig:LNM:III:en}$ (b,f,g).
and $\delta\dot{\varphi}_{1}=0.01\dot{\varphi}_{1}$. 
The more energy transfer is  in the counter-rotating oscillations than in co-rotating.

At high energy level $\left(\mathcal{H}=100.898\right)$ with initial conditions:
 $\theta=1.16073984$, $\varphi=0.0$,
$\theta_{1}=-0.32498693$, $\varphi_{1}=0.0$, $\theta_{2}=-0.32498693$,
$\varphi_{2}=0.0$, $\dot{\theta}=0.0$, $\dot{\varphi}=7.90726534$,
$\dot{\theta}_{1}=0.0$, $\dot{\varphi}_{1}=-2.75298937$, $\dot{\theta}_{2}=0.0$,
$\dot{\varphi}_{2}=-2.75298937$ (Fig. $\ref{fig:LNM:III:en}$(c,
g, k)). For counter-rotating case, the initial velocity of second pendulum
is  changed to $\dot{\varphi}_{2}=2.75298937$. 
In these both cases, co- and counter-rotating oscillations of pendulums
are chaotic for higher total energy level (Fig. $\ref{fig:LNM:III:en}$(d,
h, l)).

\section{Summary}
\label{sec:Summary}

In this paper we study the dynamics of three coupled conservative
spherical pendulums. The analytical calculations of the eigenfrequencies
of the linearized system, allow us to identify three independent linear
modes of the pendulum's rotation. For all of them, pendulums rotate
in clock-wise direction with different phase shifts. The obtained
linear modes allow us to compute the nonlinear normal modes for increasing
energy in the system using the path-following method. As it is expected
with growing total energy the frequencies of modes increase. In the
first mode we observe a pitchfork bifurcation, which causes the appearance
of symmetry broken periodic solution and destabilization of the initial one. 

In the case of periodic motion there is no energy transfer between the pendulums,
but even a small perturbation of periodic initial conditions leads
to the exchange of energy. We show the energy flows for each mode considering
two cases of rotations - in  clock-wise and counter clock-wise directions.
In the first case for all three modes we observe the  dynamics on a KAM
tori. The period of energy transfer is much longer than the natural
period of each mode. In the second case the dynamics is always chaotic.

\section*{Acknowledgments}
This work has been supported by the Foundation for Polish Science,
Team Programme under project TEAM/2010/5/5 (BW, PP, TK). AP would
like to thank the DST, Govt. of India for financial support.


\begin{thebibliography}{}



\bibitem{Pikovsky2001}
A. Pikovsky, M. Rosenblum, J. Kurths, Cambridge University Press, {2001}

\bibitem{Strogatz2000}
S.~H. Strogatz, Physica D: Nonlinear Phenomena \textbf{143}, {2000} 1

\bibitem{Kuramoto1984}
Y. Kuramoto, Springer, {1984}

\bibitem{Sabarathinam20133098}
S. Sabarathinam, K. Thamilmaran, L. Borkowski, P. Perlikowski, P. Brzeski, A. Stefanski, T. Kapitaniak, Communications in Nonlinear Science and Numerical Simulation \textbf{18}, {2013} 3098

\bibitem{Yanchuk2005c}
S. Yanchuk, K. Schneider, \textit{Proceedings of Equadiff 2003} (World Sci.,2005) 494-496

\bibitem{vakakis2008nonlinear}
A. Vakakis, O. Gendelman, L. Bergman, D. McFarland, G. Kerschen, Y. Lee, Springer \textbf{156}, {2008}

\bibitem{Saxena2012205}
G. Saxena, A. Prasad, R. Ramaswamy, Physics Reports \textbf{521}, {2012} 205

\bibitem{Wierschem2012}
N.~E. Wierschem, D.~D. Quinn, S.~A. Hubbard, M.~A. Al-Shudeifat, D.~M. McFarland, J. Luo, L.~A. Fahnestock, B.~F.~S. Jr., A.~F. Vakakis, L.~A. Bergman, Journal of Sound and Vibration \textbf{331}, {2012} 5393

\bibitem{doi:10.1142/S0218127412501003}
G. Rega, S. Lenci, B. Horton, M. Wiercigroch, E. Pavlovskaia, International Journal of Bifurcation and Chaos \textbf{22}, {2012} 1250100

\bibitem{czolczynski:023129}
K. Czolczynski, P. Perlikowski, A. Stefanski, T. Kapitaniak, Chaos: An Interdisciplinary Journal of Nonlinear Science \textbf{21}, {2011}

\bibitem{Brzeski20125347}
P. Brzeski, P. Perlikowski, S. Yanchuk, T. Kapitaniak, Journal of Sound and Vibration \textbf{331}, {2012} 5347

\bibitem{olsson1978}
M.~G. Olsson, American Journal of Physics \textbf{46}, {1978} 1118

\bibitem{olsson1981}
M.~G. Olsson, American Journal of Physics \textbf{49}, {1981} 531

\bibitem{14}
R. Cushman, J.~J. Duistermaat, Bulletin (New Series) of the American Mathematical Society \textbf{19}, {1988} 475

\bibitem{15}
V. Guillemin, A. Uribe, Communications in Mathematical Physics \textbf{122}, {1989} 563

\bibitem{Leyendecker2004}
S. Leyendecker, P. Betsch, P. Steinmann, Computational Mechanics \textbf{33}, {2004} 174

\bibitem{jp9617128}
P.~H. Richter, H.~R. Dullin, H. Waalkens, J. Wiersig, The Journal of Physical Chemistry \textbf{100}, {1996} 19124

\bibitem{2011arXiv1103.2822L}
T. Lee, M. Leok, N. Harris McClamroch, ArXiv e-prints, \textbf{arXiv:1103.2822}, {2011}

\bibitem{bender:173}
C.~M. Bender, B.~K. Berntson, D. Parker, E. Samuel, American Journal of Physics \textbf{81}, {2013} 173

\bibitem{PhysRevE.85.046117}
H.~C. Mayer, R. Krechetnikov, Phys. Rev. E \textbf{85}, {2012} 046117

\bibitem{raey}
A. Steindl, H. Troger, \textit{Bifurcation: Analysis, Algorithms, Applications} (Birkhauser Basel, 1987) 277-287

\bibitem{3}
J.~E. Marsden, J. Scheurle, Zeitschrift fur angewandte Mathematik und Physik ZAMP \textbf{44}, {1993} 17

\bibitem{doi:10.1137/040616681}
P. Chossat, N. Bou-Rabee, SIAM Journal on Applied Dynamical Systems \textbf{4}, {2005} 1140

\bibitem{11}
S. Hu, E. Leandro, M. Santoprete, Regular and Chaotic Dynamics \textbf{17}, {2012} 36

\bibitem{Shaw199385}
S. Shaw, C. Pierre, Journal of Sound and Vibration \textbf{164}, {1993} 85

\bibitem{kerschen2009nonlinear}
G. Kerschen, M. Peeters, J.-C. Golinval, A.~F. Vakakis, Mechanical Systems and Signal Processing \textbf{23}, (2009) 170

\bibitem{peeters2009nonlinear}
M. Peeters, R. Vigui{\'e}, G. S{\'e}randour, G. Kerschen, J.-C. Golinval, Mechanical systems and signal processing  \textbf{23},(2009) 195

\end{thebibliography}
\end{document}